\pdfoutput=1
\documentclass{aa}  

\newcommand*\newthanks{\unskip\hbox{\@textsuperscript{,}}}

\def\sh{Sh~2-129}
\def\hr{HR~8119}
\def\oiii{[O\,{\sc iii}]}
\def\neiii{[Ne\,{\sc iii}]}
\def\ha{H$\alpha$}
\def\hii{H\,{\sc ii}}
\def\hb{H$\beta$}
\def\hg{H$\gamma$}
\def\nii{[N\,{\sc ii}]}
\def\sii{[S\,{\sc ii}]}
\def\oiv{[O\,{\sc iv}]}
\def\oii{[O\,{\sc ii}]}
\def\oi{[O\,{\sc i}]}
\def\kms{\relax \ifmmode {\,\rm km\,s}^{-1}\else \,km\,s$^{-1}$\fi}
 \def\neiii{[Ne\,{\sc iii}]}
\def\Ne{$N_{\mathrm e}$}
\def\Te{$T_{\mathrm e}$}

\usepackage{graphicx}
\usepackage{ulem}

\usepackage{txfonts}
\usepackage[bookmarks=true,bookmarksopen=true,bookmarksopenlevel=10,bookmarksnumbered=true,colorlinks=true,linkcolor=blue,urlcolor=blue]{hyperref}
\usepackage{natbib}
\bibpunct{(}{)}{;}{a}{}{,} 
\begin{document} 

\title{Gas physical conditions and kinematics of the giant outflow Ou4
\thanks{Based on observations, obtained under Director's Discretionary
  Time of the Spanish Instituto de Astrof{\'\i} sica de Canarias, with
  the 2.5m~INT and the 4.2m~WHT telescopes operated on the island of
  La Palma by the Isaac Newton Group of Telescopes in the Spanish
  Observatorio of the Roque de Los Muchachos.}}

\titlerunning{Gas physical conditions and kinematic of the giant outflow Ou4}


   \author{Romano L.M. Corradi
          \inst{1,2}\fnmsep\thanks{{These authors equally contributed to this work.}}
          \and
          Nicolas Grosso\inst{3}{\unskip\hbox{$^{,\star\star}$}}
          \and
          Agn{\`e}s Acker\inst{3}
          \and
          Robert Greimel\inst{4}
          \and
          Patrick Guillout\inst{3}
          }
   
   \institute{
Instituto de Astrof{\'\i}sica de Canarias, E-38200 La Laguna, 
Tenerife, Spain \email{rcorradi@iac.es}
   \and
Departamento de Astrof{\'\i}sica, Universidad de La Laguna, 
E-38206 La Laguna, Tenerife, Spain 
   \and
Observatoire Astronomique de Strasbourg, Universit{\'e} de Strasbourg, CNRS, UMR 7550, 11 rue de l'Universit{\'e}, 67000 Strasbourg, France
    \and
IGAM, Institut f{\"u}r Physik, Universit{\"a}t Graz, Universit{\"a}tsplatz 5/II, A-8010 Graz, Austria\\
             }

   \date{}

 
\abstract
{Ou4 is a recently discovered bipolar outflow with a projected size of
  more than one degree in the plane of the sky. It is apparently
  centred on the young stellar cluster -- whose most massive
  representative is the triple system \hr\ -- inside the \hii\ region
  \sh. The driving source, the nature, and the distance of Ou4 are not
  known.}
{The basic properties of Ou4 and its environment are investigated in
  order to shed light on the origin of this remarkable outflow.}
{Deep narrow-band imagery of the whole nebula at arcsecond resolution
  was obtained to study its detailed morphology. Long-slit
  spectroscopy of the tips of the bipolar lobes was secured to
  determine the gas ionization mechanism, physical conditions, and
  line-of-sight velocities. An estimate of the proper motions at the
  tip of the south lobe using archival plate images is attempted.
  The existing multi-wavelength data for \sh\ and \hr\ are also
    comprehensively reviewed.}
{The observed morphology of Ou4, its emission-line spatial
  distribution, line flux ratios, and the kinematic modelling
  developed adopting a bow-shock parabolic geometry, illustrate the
  expansion of a {\sl shock-excited} fast collimated outflow.  The
  observed radial velocities of Ou4 and its reddening are consistent
  with those of \sh\ and \hr. The improved determination of the
    distance to \hr\ (composed of two B0~V and one B0.5~V stars) and
    \sh\ is 712~pc.  We identify in WISE images a $5\arcmin$-radius
  (1 pc at the distance above) bubble of emission at 22~$\mu$m
  emitted by hot (107~K) dust grains, located inside the central
  part of Ou4 and corresponding to several \oiii\ emission features of
  Ou4. }
{The apparent position of Ou4 and the properties studied in this work
  are consistent with the hypothesis that Ou4 is located inside the
  \sh\ \hii\ region, suggesting that it was launched some 90,000 yrs
  ago by \hr. The outflow total kinetic energy is estimated to
    be $\approx4\times10^{47}$~ergs. However, the alternate
  possibility that Ou4 is a bipolar planetary nebula, or the result of
  an eruptive event on a massive AGB or post--AGB star not yet
  identified, cannot be ruled out.}  
 \maketitle
%

\section{Introduction}

The remarkable nebula \object{Ou4} was discovered by the French amateur
astronomer Nicolas Outters while imaging the \sh\ nebula in June 2011
by means of a 12.5~hour CCD exposure with a F/5 106mm-diameter
refractor (resulting in a plate scale of $3\farcs5$~pix$^{-1}$) and a
narrow ($\sim$30~\AA\ FWHM) \oiii\ filter. It is a clear example of
the relevant science that can be done with very small telescopes: the
long integrations that can be obtained by dedicated amateur
astronomers allow very faint detection levels to be reached, which are
sometimes difficult to achieve with professional telescopes given the
limited amount of time available\footnote{Note that the detection
  limit per resolution element of an extended, uniform source does not
  depend on the telescope size, but only on its focal ratio.}.

Ou4 was introduced and discussed by \cite{acker12}. Its
highly bipolar morphology is mainly visible in the
\oiii~5007~\AA\ light. The total length of the nebula is $1\fdg2$ on 
the sky. 
The driving source, the nature, and the distance of Ou4 are not known. 
\cite{acker12} discussed several possibilities, including the one that
Ou4 is a very nearby planetary nebula (PN). 

Motivated by the impressive size and remarkable morphology of Ou4, we
have obtained new optical images and spectra with the aim to shed
light on its origin.
In Sect.~\ref{sh}, we describe the relative position in the sky of Ou4
and the \hii\ region \sh, and discuss the distance to the latter. 
Our imaging and spectroscopic observations are presented
in Sect.~\ref{observations}. We analyse these results in
Sect.~\ref{analysis}, and discuss the nature of Ou4 in
Sect.~\ref{discussion}.


\section{Ou4 and the \sh\ nebula}
\label{sh}

\begin{figure}[!t]
\center
\includegraphics[width=1.0\columnwidth]{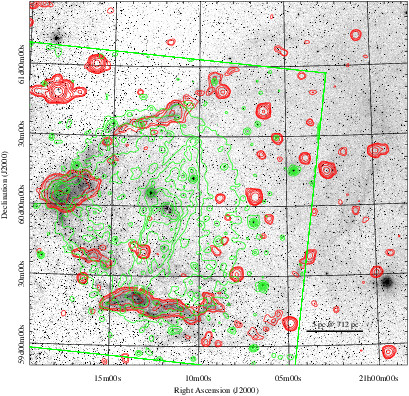}
\caption{The blister \hii\ region Sh2-129 and Ou4. The DSS2 $R$-band
  image is shown in greyscale in the background. The red contour map
  is the Green Bank 6~cm survey \citep{condon94} using 10
  contours from 0.01 to 2.8~Jy/beam with a linear step. The green
  contour map is a smoothed and thresholded \oiii\ image.  The image
  and contour map scales are logarithmic. The white arrowed line has a
  length of $1\fdg2$ and indicates the position of the giant outflow
  Ou4.  \hr, the young massive star ionising \sh, is close to the
  middle of this line.}
\label{blister}
\end{figure}

\begin{figure}[!h]
\center
\includegraphics[width=1.0\columnwidth]{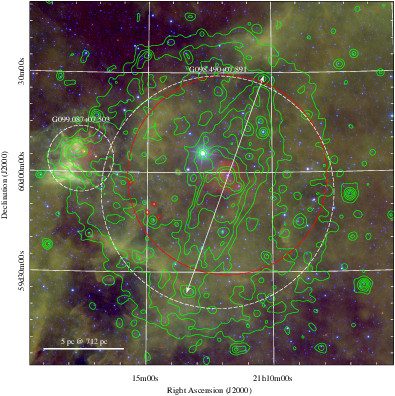}
\caption{Mid-infrared image of \sh\ obtained from the WISE All-Sky Atlas. The blue,
  green, and red colours code to the WISE infrared bands $W2$, $W3$, and $W4$ 
  centred at 4.6, 12, and 22~$\mu$m-filters, respectively. The green contour map 
  is the same \oiii\ image shown in Fig.~\ref{blister}. The dashed white circles
  indicates the two (radio quiet) \hii\ regions in the WISE catalog of
  Galactic \hii\ regions \citep{anderson14}. The red diamonds mark the
  classical T~Tauri star candidates that we have identified inside the
  30\arcmin-radius red circle centered on \hr\ (see labels in
  Fig.~\ref{2mass}).}
\label{wise}
\end{figure}

\begin{figure}[!h]
\center
\includegraphics[width=1.0\columnwidth]{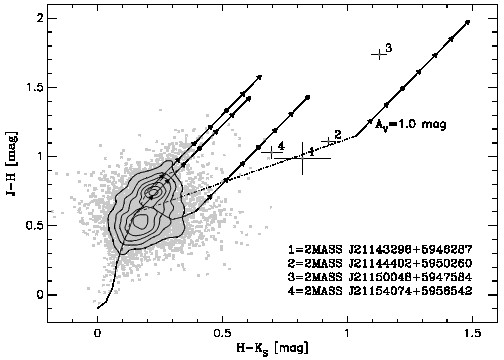}
\caption{Near-infrared colour-colour diagram of sources around
  \hr. Grey asterisks are 2MASS sources with good photometry
  located at a maximum angular distance of $30\arcmin$ from \hr\ (see
  red circle in Fig.~\ref{wise}). Contours indicate the source
  density. The continuous and dashed lines are the locii of
  main-sequence and giant stars of \cite{bessell88}.
  The dotted-dashed line is the locus of the classical T Tauri stars
  \citep{meyer97}. The arrows show the extinction vectors
  \citep{cohen81}. The classical T Tauri star candidates are labelled.}
\label{2mass}
\end{figure}

\subsection{Multi-wavelength view of the \sh\ nebula}

A multi-wavelength view of the \sh\ \hii\ region is presented in
Fig.~\ref{blister}. The $R$-band DDS2 image mainly shows the \ha\ line
emission. This defines the boundary of this ionised region, which
consists of a bright semicircle on the east side and a fainter
elongated bubble protruding towards the north-west direction. This
morphology is typical of an evolved \hii\ region in the {\sl Champagne
  phase} \citep{tenorio-tagle79}: the ionising flux of a hot young
star originally embedded near the edge of a molecular cloud produced
an expanding spherical \hii\ region which pierced the cloud surface
producing a flow of ionised gas (e.g., \citealt{yorke83}).

The Green Bank 6~cm survey (\citealt{condon94}; red contour map
in Fig.~\ref{blister}) shows the free-free continuum emission from the
border of this blister \hii\ region.  In a large field-of-view
\oiii\ image of \sh\footnote{The original image was obtained by
  another French amateur astronomer, St{\'e}phane Zoll, on August 2012
  by means of a 7.3~hour CCD-exposure with a F/3.6 106mm-diameter
  refractor (plate scale of $4\arcsec$~pix$^{-1}$) and a narrow
  ($\sim$50~\AA\ FWHM) \oiii\ filter, and posted on
  \href{http://www.astrosurf.com/ubb/Forum3/HTML/035497.html}{www.astrosurf.com}\,. We
  used {\tt Aladin} \citep{bonnarel00} to register it to the DSS2
  image.} (green contour map) the \oiii\ emission of Ou4 appears as
superposed and centred on the fainter emission from the hot gas that
fills the semi-spherical \hii\ region of \sh\ and cools within {\sl
  the Champagne flow}.

Infrared images obtained with the {\sl Wide-field Infrared Survey Explorer} 
\citep[WISE;][]{wright10b} unveil the dust emission in this area
(Fig.~\ref{wise}). Bright dust pillars are present at the eastern
border of the bubble, but they are associated to a smaller \hii\ region
according to \cite{anderson14}. This small \hii\ region and dust
pillars are typical of star-forming region. This star-forming region
may have been triggered by the expansion of \sh. In the central 
part of \sh\ a $5\arcmin$-radius bubble of emission at 22~$\mu$m shows up.
Such an emission is usually observed inside \hii\ regions where the central hot star heats 
dust grains \citep[e.g.,][]{deharveng10}; for instance, the dust pillars of 
the smaller \hii\ region are also surrounded by 22-$\mu$m emission.
The morphology of the \sh\ mid-infrared bubble and its relation with Ou4 will 
be discussed in Sect.~\ref{dust}.

We used 2MASS \citep{skrutskie06} to identify young star candidates in
the central part of \sh, in a radius of $30\arcmin$ from \hr.  We
selected only sources with good photometry (AAA flag) to build a
colour-colour near-infrared diagram in the 2MASS photometric system
(Fig.~\ref{2mass}).  The comparison with the locus of classical
T~Tauri stars (CTTSs) in this diagram\footnote{We use the colour
  transformations for the final 2MASS data release at
  \href{http://www.astro.caltech.edu/~jmc/2mass/v3/transformations/}{http://www.astro.caltech.edu/$\sim$jmc/2mass/v3/\-transformations}\,.}.
allows us to identify four CTTS candidates by their
infrared-excess. There is no source with stronger infrared-excess in
this area, i.e., no protostars.  These CTTS candidates are not located
in the vicinity of \hr, but close to the east border of our selection
area; one object has an optical extinction of about 7~mag, whereas the
other sources do not have an optical extinction larger than 1~mag.

\subsection{The distance to \hr\ and \sh}
\label{distance}

\cite{georgelin70} identified the hot (spectral type B0~V,
\citealt{hiltner56}) star \hr\ (a.k.a.\ HD~202214) as the ionising
source of \sh. Consequently, \hr\ and \sh\ must be located at the same
distance.  We also note that \hr\ is the star with the earliest
spectral type in the stellar group that is visible at the centre of
this blister \hii\ region according to {\tt SIMBAD}. This young star
is a member of the Cepheus OB2 massive stars association \citep[see,
  e.g., Fig.~3 of ][]{patel98}\footnote{Note in particular that
  \hr\ cannot belong to the Trumpler~37 stellar cluster which is
  located inside the IC~1396 \hii\ region (a.k.a.\ Sh~2-131) at an
  angular distance of $4\fdg3$ east from Sh~2-129.}.
\hr\ also lies in the central part of Ou4, as noted by \cite{acker12}.

\cite{acker12} estimated the distance of \hr\ to be 
$\sim$870~pc by assuming no foreground extinction 
and an absolute magnitude for main-sequence stars, but the distance
mostly used in the literature for \sh\ is 400~pc. This latter value is
the spectrophotometric distance derived by \cite{georgelin70} assuming
that \hr\ is a single star. However, \hr\ is a triple system, where
the separations obtained from speckle interferometry between the
primary and the Ab and B components were $0\farcs045$ and
$1\farcs021$, respectively, at epoch 2005.8654 \citep{mason09}. We
improve the distance determination by combining the dynamical mass of
{\hr}A with evolutionary tracks of massive stars with solar
metallicity \citep{bertelli94}\footnote{The {\tt CMD 2.5} web
  interface available at:
  \href{http://stev.oapd.inaf.it/cgi-bin/cmd}{http://stev.oapd.\-inaf.it/cgi-bin/cmd}
  was used.}.  The orbital elements of the {\hr}Aa and Ab components
were published by \cite{zirm12} and flagged as reliable in {\sl the
  Sixth Catalog of Orbits of Visual Binary Stars}\footnote{Catalog
  available on-line at:
  \href{http://www.usno.navy.mil/USNO/astrometry/optical-IR-prod/wds/orb6}{http://www.usno.navy.mil/USNO/astro\-metry/optical-IR-prod/wds/orb6}\,.}.
From the orbit semi-major axis ($a=0\farcs066$) and the period
($P=56.93$~yr), the following dynamical mass is obtained for {\hr}A:
$M_\mathrm{dyn}/M_\odot=32.1\times(P/\mathrm{56.93~yr})^{-2}\times(a/0\farcs066)^3\times(d/712~\mathrm{pc})^3$,
where $d$ is the distance to \hr.  The following observational
constraints are used: the spectral type B0~V for the primary
component, which corresponds to an effective temperature of 31,500~K
\citep{pecaut13}; the differential $V$-band magnitude between the Ab
and Aa components ($0.6$~mag) in {\sl The Washington Visual Double
  Star Catalog} \citep{mason01a}; and the Tycho-2 magnitudes of {\hr}A
and B, $B_\mathrm{T}=6.151$ and 6.864~mag, and $V_\mathrm{T}=6.092$
and 6.768~mag, respectively \citep{hog00}, which are converted to
$B=6.132$ and 6.838~mag, and $V=6.085$ and 6.757~mag, respectively,
using the filter transformations of \cite{mamajek02}.  Comparison of
these photometric data with the evolutionary tracks provides a
best-fit age of 3.7~Myr, which gives a total mass of 32.0~$M_\odot$
for {\hr}A\footnote{For the Aa, Ab, and B components, respectively:
  the masses are 17.8, 14.2, and 16.7~$M_\odot$; the effective
  temperatures are 31500, 29100, and 30800~K (corresponding to B0~V,
  B0.5~V, and B0~V spectral types according to \citealt{pecaut13}; 
the luminosities are 41370, 20398, and 33595~$L_\odot$; the
  intrinsic $(B-V)_0$ colours are $-0.279$, $-0.265$, and $-0.275$;
  the observed $V$-band magnitudes are 6.579, 7.179, and 6.757~mag.},
a distance of 712~pc, and a foreground visual extinction of
1.1~mag. The latter is consistent with the computed colour excess of
{\hr}B (adopting $R_\mathrm{V}=3.1$). The age determination is
consistent with the rather evolved evolutionary status of this
\hii\ region and the lack of 2MASS sources \citep{skrutskie06} with
near-infrared excess (i.e., accretion circumstellar disc) in the
vicinity of \hr.  Therefore, we adopt a distance of 712~pc
to \hr\ and \sh.  


\section{Observations}
\label{observations}

\begin{figure*}[!hp]
\centering
\includegraphics[height=23.5cm]{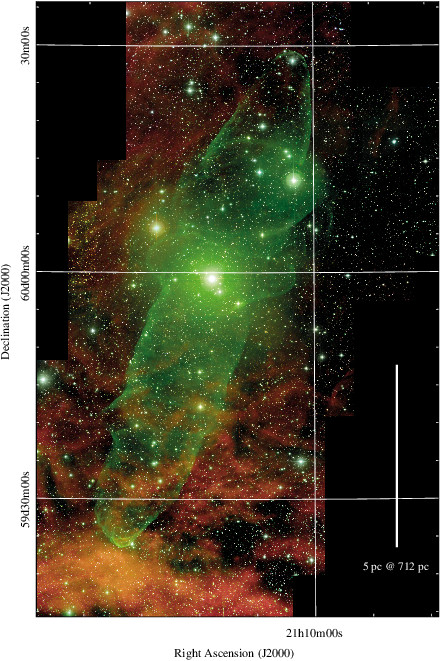}
 \caption{Colour composite mosaic image of Ou4 obtained with the Wide Field Camera of the 2.5m~INT. 
The \oiii~5007 emission is mapped in
   green, \ha+\nii\ in red, and the broadband $g$ filter in blue. Intensity scale is logarithmic. The
   field of view is $1\fdg36\times0\fdg89$. The vertical line indicates 
   the linear size if Ou4 is at the adopted distance of
   \hr\ and \sh\ (712~pc).  North is up, east is left.}
 \label{F-wholeneb}
\end{figure*}

\begin{figure*}[!ht]
\centering
\begin{tabular}{@{}c@{}c@{}c@{}}
\includegraphics[height=9.7cm]{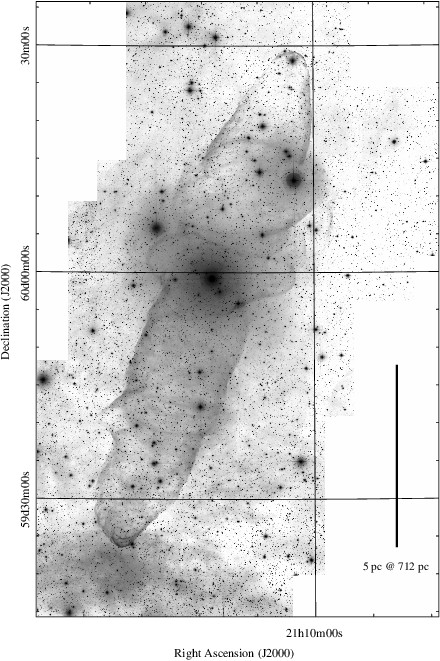}&
\includegraphics[height=9.7cm, trim = 1.2cm 0 0 0, clip=true, keepaspectratio]{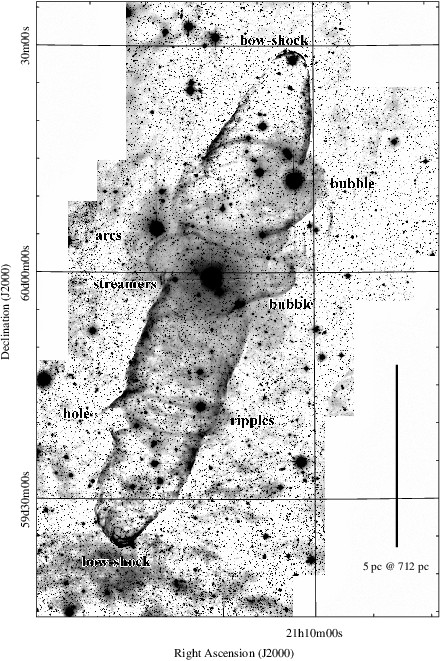}&
\includegraphics[height=9.7cm, trim = 1.2cm 0 0 0, clip=true, keepaspectratio]{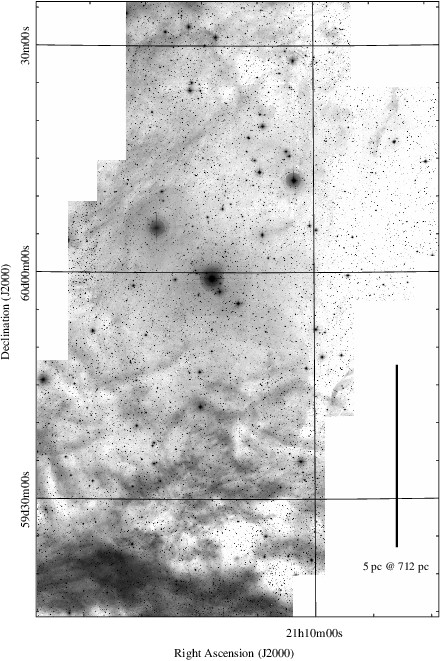}\\
\end{tabular}
\caption{The \oiii~5007 (left and middle) and \ha+\nii\ (right) mosaic
  images of Ou4.  The sharpness of the \oiii~5007 image was enhanced in the middle
    panel using a blurred mask. 
The colour scale is logarithmic in the left
    and the right panel.
The field of view in each panel is $1\degr.36\times0\degr.89$. 
North is up, east is left.}
\label{F-oiiiha}
\end{figure*}

\subsection{Imaging of Ou4}

The imaging was carried out on 18 August 2012 at the F/3.3 2.5m Isaac Newton
Telescope (INT) at the Observatorio del Roque de los Muchachos (ORM),
La Palma, Spain. Images were taken with the Wide Field Camera (WFC) in
three filters: a narrow-band \oiii\ filter, with a central wavelength
of 5008~\AA\ and a bandpass of 100~\AA; a narrow-band \ha\ filter, 
with a central wavelength of 6568~\AA\ and a bandpass of 95~\AA, 
thus including the \nii\ doublet around \ha; and a
broadband Sloan $g$ filter, with a central wavelength of 4846~\AA\ and
a bandpass of 1285~\AA.
The plate scale is $0\farcs33$ pix$^{-1}$, and the seeing varied from
$1\farcs2$ to $2\farcs0$ during of the night.  As the field of view of
the WFC is 34$\times$34~arcmin$^2$, in order to cover the totality of
Ou4 and the gaps between CCDs in the detector mosaic, six
different telescope offsets, with generous overlapping among them, were
adopted: at each of them, we exposed for a total of between 15 min and
45 min in \oiii, 5 and 15 min in \ha, and 90 sec in the $g$
filter. The night was dark: any significant moonlight would prevent
detection of the nebula over the background given its very low surface
brightness (see Sect.~\ref{analysis}).

Images were reduced using the INT+WFC pipeline in Cambridge
(CASU). Precise astrometric solution allowed a careful combination of
all images using the {\it swarp} software \citep{bertin02}. The
final mosaic has a field of view of roughly $0\fdg9\times1\fdg4$. As
Ou4 and the surrounding \hii\ region \sh\ cover the whole field of
view of the camera, accurate background subtraction proved to be
difficult and had to be carefully tuned.
Figure~\ref{F-wholeneb} shows the colour-composite INT image of Ou4,
while Fig.~\ref{F-oiiiha} shows the \oiii\ and \ha+\nii\ images with
labels on the most relevant morphological features discussed
below. 

\subsection{Spectroscopy of the lobe tips}
\label{spectroscopy}

Spectra of Ou4 were obtained on 17 August 2012 with the 4.2m~WHT
telescope and the double-arm ISIS spectrograph. The long slit of ISIS
was opened to 1\arcsec-width and positioned in the southern tip of the
nebula at P.A.=344$\degr$, crossing the bright ridge of the nebula at coordinates
R.A.=$21^{\mathrm h}13^{\mathrm m}23\fs8$ and Dec=$+59\degr23\arcmin40\farcs5$ (J2000.0),  
as indicated in Fig.~\ref{F-detneb} by the long blue slit. 

In the blue arm of ISIS, grating
R300B was used, providing a dispersion of 1.7~\AA\ per (binned
$\times$2) pixel, a resolution of 3.5~\AA, and a spectral coverage
from 3600 to 5200~\AA. In the red arm, grating R158R gave a dispersion
of 1.84~\AA~pix$^{-1}$, a resolution of 3.5~\AA, and a spectral
coverage from 5400 to 7800~\AA.
Total exposure times was 2 hours.
The spatial scale was $0\farcs4$ per binned pixel in the blue, and $0\farcs44$ in the
red. Seeing was $1\farcs0$.  The spectrophotometric standard BD+28~4211
from \cite{oke90} was observed during the night for flux calibration.

In order to determine the basic kinematic properties of the outflow,
spectra at a higher resolution were also obtained on 24 October and 25
December 2012 using the same instrument but grating 1200B.  Useful
data are limited to the \oiii~5007~\AA\ line, at which a spectral
dispersion of 0.22~\AA~pix$^{-1}$ and a resolution of 0.84~\AA\ was
achieved with the adopted $0\farcs9$ slit width.  The slits, also
indicated in Fig.~\ref{F-detneb} in red, cross the south and north
tips of the lobes of Ou4 at positions $21^{\mathrm h}13^{\mathrm
  m}25\fs2$, $+59\degr23\arcmin46\farcs7$ (J2000) at P.A.=340$\degr$, and
$21^{\mathrm h}10^{\mathrm m}21\fs8$, $+60\degr29\arcmin10\farcs6$ at
P.A.=20$\degr$, respectively.  Total exposure times were 40 min in the
south lobe, and 80~min in the fainter north lobe.

Spectra were reduced with the standard procedure using the {\it
  longslit} package of {\it iraf V2.16}\footnote{Iraf is distributed by the
National Optical Astronomy Observatory, which is operated by the
Association of Universities for Research in Astronomy (AURA) under
cooperative agreement with the National Science Foundation.}.

\section{Analysis}
\label{analysis}

\subsection{Overall morphology of Ou4}
\label{morphology}

The bipolar nebula is mainly visible in the \oiii~5007 light (green
colour in Fig.~\ref{F-wholeneb}, and left and middle panels of
Fig.~\ref{F-oiiiha}), whose surface brightness ranges from a few
10$^{-16}$~erg~cm$^{-2}$~s$^{-1}$~arcsec$^{-2}$ (tip of southern lobe,
see Sect.~\ref{gas}) down to the detection limit of our images of
several 10$^{-17}$~erg~cm$^{-2}$~s$^{-1}$~arcsec$^{-2}$ (in regions
not contaminated by the diffraction halo of bright stars).

\begin{figure}[!t]
\centering
\includegraphics[width=\columnwidth]{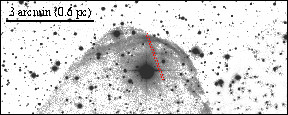}\\[3pt]
\includegraphics[width=1.026\columnwidth]{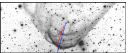}
\caption{Details of the north (top) and south (bottom) bow-shocks in
  the \oiii\ light. The field of view in each panel is
  $10\arcmin\times4\arcmin$. North is up, east is left.  The slit
  position for the lower resolution spectrum is indicated by the long
  (blue) slit. The reference star 2MASS~J21131972+5925365
  adopted as the zero--point of the $x$--axis in Fig.~\ref{F-lprof} is 
visible at its north end. The positions for the
  higher resolution spectra are indicated by the short (red) slits.}
\label{F-detneb}
\centering
\includegraphics[width=\columnwidth]{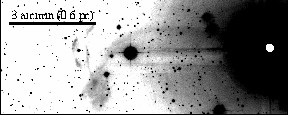}
\caption{The innermost region of the south lobe in the \oiii\ light.
  The field of view is $10\arcmin\times4\arcmin$. North is up, east is
  left.  Its eastern edge has a marked curvature and approximately
  points toward \hr\ (indicated by the white circle).}
 \label{F-cenlobe}
\end{figure}

As described in \cite{acker12}, Ou4 is mainly composed of two
collimated lobes with arc-shaped tips of enhanced \oiii\ emission.
Their extremities recall the bow-shocks observed in collimated stellar
outflows such as Herbig-Haro objects in the outflows from Young
Stellar objects \citep{reipurth01}, or in high velocity bipolar PNe.
The south lobe, of cylindrical shape, is longer than the northern one,
which instead has a more conical overall morphology. Neither lobe can
be easily followed back to the central region, where a large distorted
``bubble'' is visible.  This prevents a safe association of the nebula
with one of the stars near its symmetry centre. Only the eastern ridge
of the south lobe extends back enough to provide some useful
indication. The lobe is wide till close to centre, where its edge is
bowl shaped and bends toward a direction approximately pointing to
\hr\ (Fig.~\ref{F-cenlobe}). However, detection of the faint emission
from the lobe at distances smaller than $\sim2.2\arcmin$ from \hr\ are
prevented by the bright diffraction halo of this star.
This marked lobe curvature near the centre is similar, for instance,
to that of the inner lobes of the bipolar PN Mz~3
\citep{santander-garcia04} or of several proto-PN such as Hen~401
\citep{sahai99,balick13}.  The overall morphology of Ou4 is very similar to 
that of the giant bipolar PN KjPn~8 \citep{lopez95}, which has a 
size of $7.3$~pc $\times$ $2.1$~pc for the kinematically-determined distance of
$1.8\pm0.3$~kpc \citep{boumis13}.

Compared to the discovery image, the INT mosaic reveals a wealth of
new details. In addition to the above-mentioned central bubble,
another elliptical bubble breaks the inner regions of the northern
lobe. Also, in the central region of the south lobe a ``hole'' is
visible as well as ``ripples'' directed perpendicularly to its
long-axis. East of the central distorted bubble, additional features
in the form of ``streamers'' and ``arcs'' can be identified
(Fig.~\ref{F-oiiiha}).
Finally, the tips of the lobes, and in particular the southern one,
are composed of complex systems of multiple arcs 
(see also Fig.~\ref{F-detneb}). 
The main properties of Ou4 determined in this work are summarised
  in Table~\ref{T-properties}.

In the \ha+\nii\ filter (red colour in Fig.~\ref{F-wholeneb}, and
right panel of Fig.~\ref{F-oiiiha}), emission is dominated by ambient
gas belonging to the \hii\ region \sh\ with little contribution from
Ou4 except at specific regions like the south and north bow-shocks.

\begin{table}[!ht]
\caption{The main properties of Ou4.}
\begin{tabular}{lll@{}}       
\hline\hline
\multicolumn{3}{c}{{Ou4}} \\ 
\hline

{Lobe angular size}           &  $1\fdg16\times{0\fdg20}$ & \\
{Linear size}           &  {14.4~pc $\times$ 2.5~pc} & {(712 pc)}\\
Centre$^\ast$  & $21^{\mathrm h}11^{\mathrm m}45.6$ $+59\degr59\arcmin00\arcsec\phantom{.0}$ & Equ.\ J2000\\
\hr            & $21^{\mathrm h}11^{\mathrm m}48.2$ $+59\degr59\arcmin11\farcs8$      & Equ.\ J2000\\
               & {$98\fdg5202\phantom{000}$ $+07\fdg9852$}              & {Gal.\ J2000}\\
\hline
\multicolumn{3}{c}{{Tip of the South lobe}} \\ 
\hline
Optical extinction        &  1.1$\pm$0.4~mag & \\
{\Te(\oiii,~\oii)}         &  {55,000~K, $\ge$20,000~K} & See {\S\ref{gas}} \\
\Ne(\sii)         &  {$\ge$}50--100~cm$^{-3}$                 & See {\S\ref{gas}} \\
Proper motion     &  ${\lesssim}$$0\farcs06$$\pm$$0\farcs03$~yr$^{-1}$ tentative & {See \S\ref{proper_motion}}\\
RV range$^\dag$ &  $-100$~---~$+10$~\kms &          \\
{Velocity} & {112~\kms} & {See \S\ref{model}}\\
\hline
\multicolumn{3}{c}{{Tip of the North lobe}} \\ 
\hline
RV range$^\dag$ &  $-55$~---~$+40$~\kms &             \\
{Velocity} & {83~\kms} & {See \S\ref{model}}\\
\hline
\end{tabular}
\newline $^\ast$ Approximate symmetry centre of the Ou4 nebula.
\newline $^\dag$ Heliocentric, determined in the region covered by the slit (see Fig.~\ref{F-vel}).
\label{T-properties}
\end{table}

\subsection{Emission line distribution in the tip of the south lobe}

Insights into the excitation and dynamics of the nebula are gained by
looking at the spatial distribution of different ions. This is shown
in Fig.~\ref{F-lprof} for the lower resolution ISIS spectrum cutting
the tip of the southern lobe approximately south-north (cf.\ bottom
panel of Fig.~\ref{F-detneb}).  The upper panel of Fig.~\ref{F-lprof}
shows the spatial distribution of representative lines with the
highest ionisation potential, such as \oiii~5007 and \neiii~3869,
while the middle and bottom panels present those with lower ionisation
potential (\ha, \nii, \sii, and \oii).  Profiles were smoothed with a
boxcar filter size of 3 spatial pixels ($1\farcs3$). The field star
2MASS~J21131972+5925365, visible at the upper end of the long
blue slit in the bottom panel of Fig.~\ref{F-detneb}, was used to
anchor the blue and red spectra and define the zero--point of the
spatial scale. Negative distances refer to positions south of the
reference star.

The sharp edge of the southern lobe corresponds to the peak in the
\oiii~5007 emission at a distance $d=-115\arcsec$.  \neiii\ has a
similar behaviour, though the main peak is not equally pronounced.
Ahead of the \oiii\ peak (i.e., at $d<-115\arcsec$), significant
\ha\ and \nii, and to less extent \oii\ and \sii\ emission, can be
associated with the \hii\ region \sh. These lower ionisation ions also
contribute to the emission of Ou4, as shown by the several peaks at
$d>-115\arcsec$, but their overall distribution is obviously shifted
to the north (i.e., ``inside'' the lobe) compared to \oiii. In
particular, \oii~3727, which is at least as bright as \oiii~5007 (even
taking into account mixing with the emission of \sh), peaks about two
arcseconds north of \oiii.  Some faint \oi~6300 emission, not shown in
Fig.~\ref{F-lprof}, is also detected inside the lobe.  No significant
differences are seen in the profiles of other lines not shown in
Fig.~\ref{F-lprof}, and in particular \oiii~4363 is similar to
\oiii~5007, \hb\ to \ha, and \sii~6731 to \sii~6716.

This overall distribution, with the higher ionisation ions outward and
the lower ionisation ones progressively inward, is a first indication
that the gas in the tip of the south lobe of Ou4 is
shock-excited. Photoionisation from a central source would instead
produce an opposite stratification with the higher ionisation ions
inwards.  

\subsection{Gas physical conditions from emission line ratios}
\label{gas}

\begin{figure*}
\centering
\includegraphics[width=13.0cm]{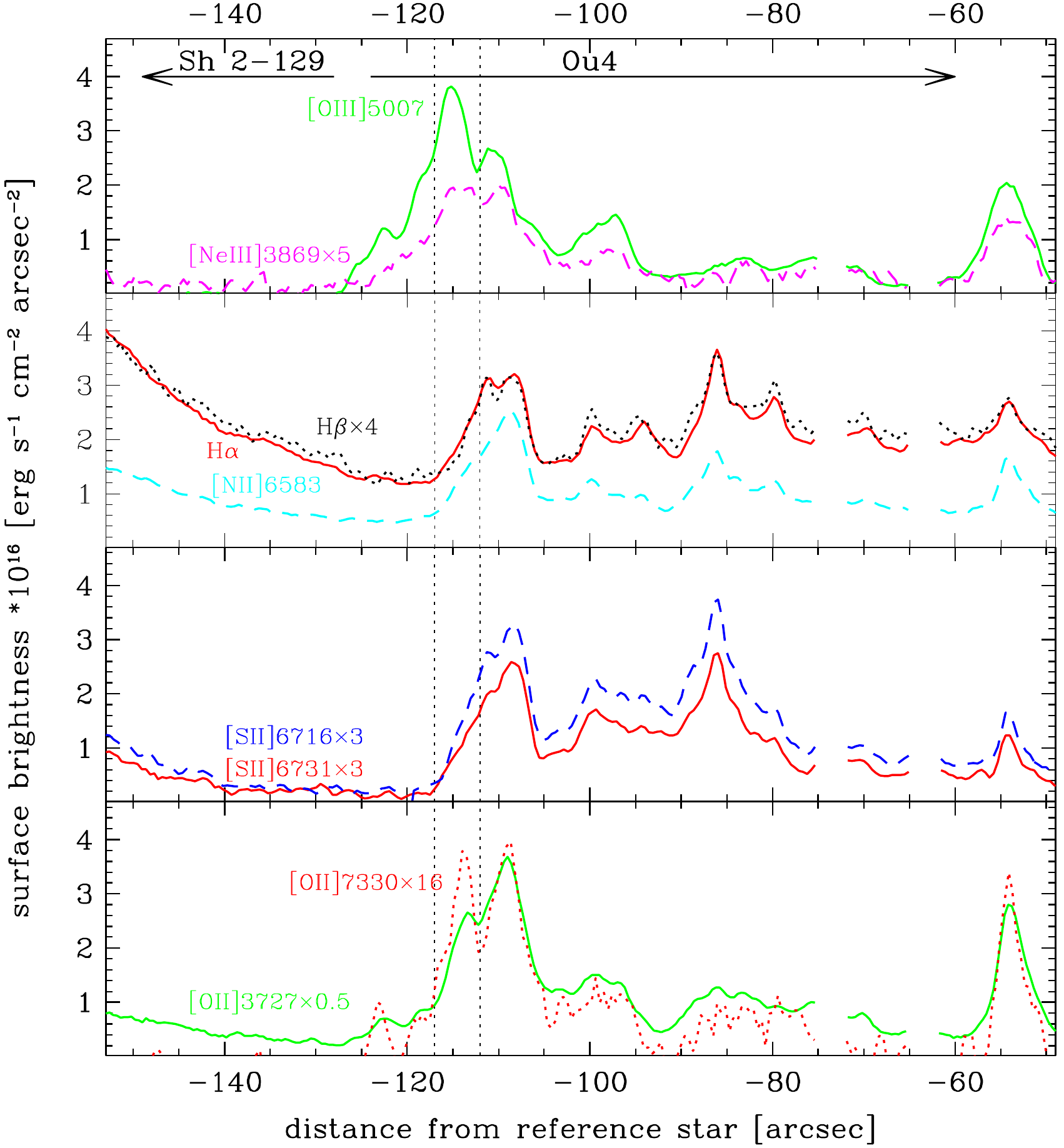}
 \caption{Spatial profile of selected emission lines through the tip
   of the southern lobe, along the slit shown in blue colour in
   Fig~\ref{F-detneb}. Emission of different ions has been scaled
   by the factors indicated in the labels. Regions corresponding to
   field stars are masked. The vertical dotted line show the
   integration limits for the line flux measurements presented in
   Table~\ref{T-linefluxes}.}
 \label{F-lprof}
\end{figure*}

Line fluxes for the Ou4 outflow were measured in the low-resolution
spectrum crossing the tip of south lobe. A precise background
subtraction is limited by the overlapping emission of the \hii\ region
\sh, which surface brightness is highly variable at different spatial
scales as it can be seen in the right panel of Fig.~\ref{F-oiiiha} and
in Fig.~\ref{F-wholeneb}.  We have done several tests, and finally
adopted a background (night sky + \sh) determined using the portion of
the spectrum of the northernmost part of the slit, near the adopted
reference star (Fig.~\ref{F-detneb}). This is likely to leave some
small contamination by \sh. Note that, however, the main conclusions
of the analysis presented below are not affected by a different choice
of the region used for background subtraction.

The ionisation stratification discussed in the previous section makes
the line ratios highly dependent on the selected spatial
region over which flux is integrated. We present in
Table~\ref{T-linefluxes} line fluxes measured in the shocked gas
between $-117$\arcsec\ and $-112$\arcsec\ from the reference star in the south
lobe, around the main peak of the \oiii\ emission. These integration
limits are indicated by the vertical dotted lines in
Fig.~\ref{F-lprof}, and do not include the cooling region of lower
excitation behind the shock.

In order to derive physical quantities from these fluxes, the nebular
extinction is required.  The logarithmic extinction constant
  $c_\beta$ has been determined from the Balmer decrement by averaging
  the results from the observed \ha/\hb\ and \hg/\hb\ ratios in our
  long-slit spectra. The theoretical Balmer line ratios were adopted 
  from \cite{brocklehurst71} for electron temperatures between 10000 and 20000~K
  and a low density regime ($N_e\sim10^2$, see below). Rather than
  using only the narrow spatial region of the shocked gas around the
  \oiii\ peak, to determine an average value of the extinction of the
  south lobe of Ou4 (which can be patchy at these relatively large
  angular sizes) we have considered a broader area between
  $-117$\arcsec\ and $-105$\arcsec\ from the reference star, where
  most of the Balmer line emission in the region covered by our long
  slit is produced (see Fig.~\ref{F-lprof}). This also includes part
  of the cooling region behind the shock. We derive a value of
  $c_\beta$=0.5$\pm$0.2 using the reddening law of 
  {\cite{fitzpatrick99}}
  and $R_\mathrm{V}$$=$3.1. This corresponds to
  $A_\mathrm{V}$=1.1$\pm$0.4~mag. The same value, within the errors,
is found for the emission from \sh\ ahead of the shock. This $c_\beta$
value was adopted to deredden the fluxes in Table~\ref{T-linefluxes}.
This foreground extinction value is also similar to the one that we
found for \hr\ in Sect.~\ref{sh}.

\begin{table}[!t]
\caption{Observed and dereddened (adopting $c_\beta$=0.5) line 
fluxes in the shock
  region between -112 and -117\arcsec\ (see
  Fig.~\ref{F-lprof}) of the south lobe. Fluxes are normalised
  to H$\beta$=100. The observed H$\beta$ integrated flux in the area is 4.03\,
  10$^{-16}$~erg~cm$^{-2}$~s$^{-1}$. The derived physical
    quantities are \Ne(\sii)=50--100~cm$^{-3}$ and \Te(\oiii)=55,000~K.}
\centering
\begin{tabular}{lrr}       
\hline\hline\\[-7pt]                
Line identification & \multicolumn{2}{c}{Flux} \\
                    & Observed & Dereddened    \\
\hline\\[-5pt]                     
\oii~3726+3729          &    883.1 &  1262.3\\ 
\neiii~3869             &     72.5 &    99.3\\ 
\neiii~3968+H$\epsilon$ &     35.9 &    47.7\\ 
H$\delta$   ~4101       &     18.9 &   24.1  \\ 
H$\gamma$   ~4340       &     45.7 &   53.8  \\ 
\oiii       ~4363       &     55.4 &   64.7  \\ 
H$\beta$    ~4861       &    100.0 &  100.0  \\ 
\oiii       ~4959       &    237.8 &  231.1  \\ 
\oiii       ~5007       &    717.5 &  687.5  \\ 
\nii        ~6548       &     94.3 &   63.1  \\ 
H$\alpha$   ~6563       &    503.7 &  336.1  \\ 
\nii        ~6583       &    308.3 &  205.0  \\ 
\sii        ~6716       &    114.4 &   74.2  \\ 
\sii        ~6731       &     87.0 &   56.3  \\ 
\oii        ~7319       &     51.8 &   30.6  \\ 
\oii        ~7330       &     53.7 &   31.7  \\ 
\hline          
\end{tabular}
\label{T-linefluxes}
\end{table}

Physical conditions in the gas (electron density and temperature) were
then computed using the {\it nebular} package within {\it iraf}, based
on the five-level atom program originally published by
\cite{de_robertis87}, and further developed by \cite{shaw95}.  From
the \sii\ doublet, we obtain an electron density \Ne\ of between
50~cm$^{-3}$ and 100~cm$^{-3}$ depending on the adopted
temperature. This is a {\sl lower} limit for the gas density in the
tip of the south lobe of Ou4, as the \sii\ emission mainly comes
  from the inside boundary of the selected region of the spectrum (see
  Fig.~\ref{F-lprof}) and therefore traces a different nebular zone
  than \oiii. In addition, non-negligible residual contribution of
  \sii\ emission from the overlapping low-density \sh\ could be
  present. The electron temperature \Te\ derived from the
\oiii(5007+4959)/4363 line ratio is as large as 55,000~K for this
density. This is a very high \Te, which could be reduced to values
typical of photoionised nebulae only if \Ne\ had the value of
10$^7$~cm$^{-3}$ in the O$^{2+}$ emitting region. On the other hand,
\oii\ is expected to be produced in the same region as \sii\ (but
  note the additional emission peak of \oii7330 in
  Fig.~\ref{F-lprof}), and therefore the \sii\ density can be adopted
to compute \Te\ from the \oii(3726+3729)/(7320+7330) line ratio. The
latter ratio also suggests a high electron temperatures
($\ge$20,000~K), even though at low densities the \oii\ indicator runs
into its asymptotic value.

Such a \Te, too high to be produced by photoionisation, is another
proof that the gas at the tip of the lobes of Ou4 is ionised by
shocks.  This conclusion is further supported by standard shock
excitation diagnostic line ratios. In this nebular region,
$\log$(\ha/\nii6548+6583)$=$$0.1$ and
$\log$(\ha/\sii6716+6731)$=$$0.4$, which locates Ou4 together with
shock-excited sources such as Herbig--Haro objects (see e.g.\
\citealt{viironen09}). 
The same conclusion is reached by analysing other
portions of the slit.  The lower-excitation emission inside the lobes
is therefore expected to be the recombining/cooling region behind the
shock. Integrating the \nii(6548+6583)/5755 between slit positions
-117\arcsec\ and -105\arcsec, including the main \nii\ emission peak, we
consistently obtain a lower temperature, \Te$\sim$14,400~K.

The electron density of gas in the relatively bright region of
\sh\ ahead of the bow-shock (from -150\arcsec\ to
  -130\arcsec\ from reference star), determined from the
\sii6716,6731 doublet, is \Ne=55~cm$^{-3}$ assuming a typical \Te\ for
\hii\ regions of 10,000~K.  
Because this emission region is located on the ionized boundary
  of the \hii\ region, this electronic density is the one of the
  molecular cloud.

To estimate the electronic density inside the \hii\ region we use
  the \ha\ emission measure along the line-of-sight,
  $EM=\int_\ion{H}{ii} N_\mathrm{e}^2 ds$, which can be computed from
  the \ha\ surface brightness corrected from extinction, $I$, with the
  following formula:
  $EM/\mathrm{(1~cm^{-6}~pc)}=2.75\times(T/\mathrm{10,000~K})^{0.9}\times(I/1~R)$,
  which is valid for temperature between 5,000 and 10,000 K, and where
  $R$ is one Rayleigh, i.e.,
  $2.42\times10^{-7}$~ergs~cm$^{-2}$~s$^{-1}$~sr$^{-1}$ at the
  \ha\ wavelength \citep{reynolds77}. We assume for the \hii\ region a
  typical temperature of 10,000~K.  From the Virginia Tech Spectral
  line Survey (VTSS; \citealt{finkbeiner03}), which mapped the
  north-west half-part of \sh\ in \ha\ with $6\arcmin$ (FWHM)
  resolution, we obtain the radial profile from \hr\ for a position
  angle of $\approx$$26\degr$ of the \ha\ intensity in $R$ unit, and
  deredden it using $A_\mathrm{V}$=1.1~mag and the extinction law of
  \cite{fitzpatrick99}.  This intensity is modeled with a uniform
  intensity of 22.3~$R$ plus a constant-density (hemi-)spherical shell
  geometry to reproduce the limb brightening \citep[e.g.,][]{naze02},
  and convolved with a $6\arcmin$-FWHM Gaussian.  To reproduce the
  external profile we need a shell-radius of $0\fdg76$ (i.e., 9.4~pc
  at a distance of 712~pc), a shell-thickness of $0\fdg14$, and a
  shell-density of 5.1~cm$^{-3}$.  The constant-density inside the
  shell is estimated to about 1.7~cm$^{-3}$.  This lower electronic
  density of the blister \hii\ region compared to the molecular cloud
  is the result of the ionization and heating by the central
  early-type B stars, which have produced the Champagne flow.

If Ou4 is embedded in \sh, the pre-shock gas density is the electronic
density inside this \hii\ region.  Since this shock is radiative, the
gas behind the shock relaxes to the \hii\ region temperature and if
there is no magnetic field the shock compression ratio scales as
Mach-number squared \citep{draine93}. Taking $\gamma=5/3$ for the
ratio of specific heat and $\mu=0.7$ for the mean molecular weight
inside the \hii\ region, the sound velocity is 14~\kms\ and a shock
velocity of 112~\kms\ (see below Sect.~\ref{model}) gives a Mach
number of 8.0, which leads to a compression ratio of 64.  Therefore,
the post-shock gas density is estimated to 110~cm$^{-3}$.

\subsection{Kinematics of the lobe tips}
\subsubsection{Radial velocities}
\label{radialvel}

\begin{figure}
\centering
\includegraphics[width=\columnwidth]{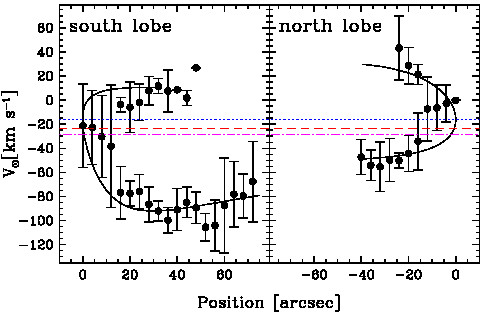}
 \caption{Position-velocity plot at the tips of the lobes. Position
   zero is defined at the point at which the slit crosses the end of
   each lobe. Positive and negative positions indicate distances from
   this reference point along the slits and toward the centre of the
   nebulae in the south and north lobe, respectively. ``Errorbars''
   indicate the \oiii\ full-width-at-half maximum after correcting for
   the instrumental broadening. Solid lines show our bow-shock
   kinematic model (see Sect.~\ref{model}). The horizontal blue dotted line
   indicate the heliocentric radial velocity of \hr, and the dashed
   and dashed-dotted lines that of \sh\ (see Sect.~\ref{radialvel}).}
 \label{F-vel}
\end{figure}

Heliocentric line-of-sight velocities in regions at the tip of each
lobe were computed from the Doppler shift of the \oiii~5007 line in
the higher resolution spectra described in Sect.~\ref{spectroscopy}.
The slit locations are indicated in Fig.~\ref{F-detneb}.  Wherever
possible, two Gaussians were fitted to the spectra at different slit
positions, binning the information every 4 arcseconds to increase the
S/N. The resulting position velocity plot is presented in
Fig.~\ref{F-vel}. Accuracy in the double-Gaussian fitting is mainly
limited by the separation and shape of the profiles of the two line
components. At the positions where they cannot be resolved, only a
single Gaussian could be measured with generally large FWHM resulting
by the blending of the two components. For this reason, we indicate in
Fig.~\ref{F-vel} the FWHM of the Gaussian line fits rather than the
(smaller) formal error of the fitting.

The plot shows that there is only a small overall velocity difference
between the two bow-shocks, indicating that the outflow is likely to
be oriented near the plane of the sky.  The north lobe would recede
from us, and the south one approach us.  Right inside each lobe, the
\oiii\ emission can be split into two components which are separated
by up to about 100\kms\ in the small fraction of the lobe length
covered by these observations. Therefore a lower limit for the lobe
``transverse'' velocity can be set to 50\kms.

In Fig.~\ref{F-vel}, the radial velocities of \hr\ and \sh\ are also
shown.  The former ($-16.2$~\kms, blue dotted line) is from
\cite{wilson53}. The \ha\ average velocity of
\sh\ ($-23.5\pm5.3$~\kms) obtained using Perot-Fabry measurements by
\cite{georgelin70} is indicated by the dashed red line, while the
dashed-dotted magenta line represents the CO peak located east of
\hr\ on the border of \sh\ ($-28.5\pm0.7$~\kms; \citealt{blitz82}).
The velocity of the two lobes is symmetrically located with respect to
the massive star and the \hii\ region, further supporting their
possible association with Ou4.

\subsubsection{Proper motions}
\label{proper_motion}

\begin{figure}[!t]
\center
\includegraphics[width=\columnwidth, trim = 0 1.1cm 0 0, clip=true, keepaspectratio]{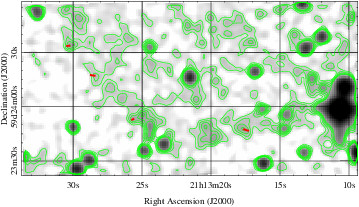}\\
\includegraphics[width=\columnwidth, trim = 0 1.1cm 0 0, clip=true, keepaspectratio]{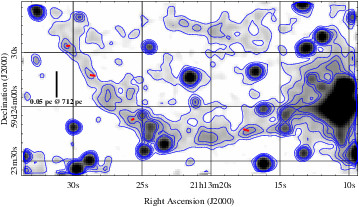}\\
\includegraphics[width=\columnwidth]{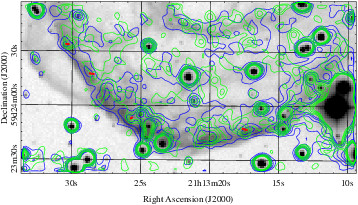}\\
\caption{
Proper motions of the tip of the south lobe of Ou4. Top and middle
  panels: POSS1 (epoch 1952-07-22) and POSS2 (epoch 1993-06-25) blue
  plates in logarithmic greyscale smoothed with a Gaussian kernel
  ($\sigma=1\farcs5$) with green and blue contours overlaid,
  respectively. Bottom panel: the green and blue 
  contours overlaid on our \oiii\ INT mosaic (epoch
  2012-08-19). The possible
  motions of four emission features  between the POSS1 and POSS2 epochs are
  indicated in each panel by small red lines (average size of $\sim
  2\farcs5\pm1\farcs1$).}
\label{pm}
\end{figure}

We identify the tip of the south lobe in
two $B$-band digitised plates obtained at two different epochs
separated by nearly 41 years (Fig.~\ref{pm}).
We use the {\tt
  imwcs} software\footnote{The {\tt
  imwcs} software is available at:
\href{http://tdc-www.harvard.edu/wcstools/imwcs}{http://tdc-www.harvard.edu/\-wcstools/imwcs}\,.}
and the {\tt SExtractor} software \citep{bertin96}
to register these $B$-band POSS1 and POSS2 images of Ou4 and our
\oiii\ INT image to
the 2MASS reference frame \citep{skrutskie06}.

However, the modest spatial resolution and the weak S/N ratio prevent
us to safely track any emission features between the POSS1 and POSS2
epochs.  Our {\sl tentative} pairing of a few emission features leads
to a proper motion upper-limit of
$\sim$$0\farcs06\pm0\farcs03$~yr$^{-1}$ (or a tangential velocity of
$\sim200\pm100$~\kms\ if Ou4 is located at the distance of \hr),
  where the proper motion error includes the error of the image
  registration and the determination of the features position.

\subsubsection{Bow-shock kinematic model}
\label{model}

The edges of the brightest working-surfaces of the south- and
north-lobe tips (Fig.~\ref{F-detneb}) can be fitted with a parabola.
Assuming axisymmetry (see Fig.~1 of \citealt{hartigan87}), we can then
argue that the intrinsic shape of the working surface is a paraboloid,
as the projection onto the plane of the sky of such a geometric figure
for any inclination angle is a parabola (see appendix of
\citealt{hartigan90}). The actual 3D shape can then be computed for a
given inclination of the bow shock along the line-of-sight.

In the shock rest frame, the parallel component of the velocity of the
incoming gas does not change through the working surface, whereas the
perpendicular component of the velocity is strongly decreased (by a
factor of 4 by the shock and about 10 by cooling; see
\citealt{hartigan87}). Therefore, in the observer frame the velocity
of the emitting post-shock material, which fills a rather thin shell
behind the working surface, is nearly perpendicular to the
paraboloid surface, its intensity is maximum at the paraboloid vertex
(i.e., equal to the shock velocity) and decreases from this
position. The observed bow-shock size is directly controlled by the
threshold value of the perpendicular component of the velocity that is
required to produce the line emission conditions.
For high inclinations the radial velocities derived by long-slit
spectroscopy have a characteristic "hook" shape in a velocity-position
diagram, where the observed range of radial velocities is of the order
of the actual bow-shock velocity (see Fig.~6 of \citealt{hartigan90}).

We build a toy model of this paraboloid bow-shock with the shape and
the size measured from our images where we can vary the systemic
radial velocity, the bow-show velocity and inclination along the line
of sight, and compute the resulting radial velocities of the emitting
post-shock material along our long-slit positions.  The systemic
radial velocity is estimated by averaging the first four (unresolved)
radial velocity measurements at the bow-shock heads, and is fixed to
this value ($-16.2$~\kms) in our simulations. We also assume the same
minimum perpendicular component of the velocity of the incoming gas in
the shock rest-frame. These simulated velocity-position diagrams are
then compared with the observed gas kinematic.

We show in Fig.~\ref{F-vel} our best match for the south and north
bow-shocks which is obtained for an inclination of $60\degr$ and
$100\degr$ (where $0\degr$ is a bow-shock directed towards us),
and a shock velocity of 112 and 83~\kms, respectively. The lower
velocity of the north bow-shock may explain why it is fainter than 
the south one. This toy models allows us to reproduce the global
properties of the observed gas kinematic.

With these shock velocities the minimum value of the perpendicular
component of the velocity of the incoming gas in the shock rest-frame
is 55~\kms. The minimum perpendicular velocity needed to observe
\oiii\ is usually supposed to be equal to 90--100~\kms, based on
plane-parallel simulation of shock emission (e.g.,
\citealt{hartigan87}). However, if Ou4 is located inside the
\sh\ \hii\ region the pre-shocked gas already contains O$^{+2}$ (see
Fig.~\ref{blister}), and the perpendicular velocity needed to produce
bright \oiii\ may be lower. Confirming this hypothesis would require
detailed emission modelling of the bow-shock and the \hii\ region
which is well beyond the scope of this article.

These south and north bow-shock velocities correspond to proper
motions of $0\farcs028\times$ and
$0\farcs024\times(d/712~\mathrm{pc})^{-1}$~yr$^{-1}$, respectively,
and to kinematical time scales of $88,300\times$ and
$87,800\times(d/712~\mathrm{pc})$~yr, respectively.  The poorly
constrained proper motion of the south bow shock obtained in the
previous section prevent us to derive a secure kinematic distance to
Ou4. However, we can likely exclude any distance lower than about
133~pc which should lead to an apparent expansion of the lobe tips
larger than our proper motion estimate plus 3 times our uncertainty.

\begin{figure}
\includegraphics[width=\columnwidth]{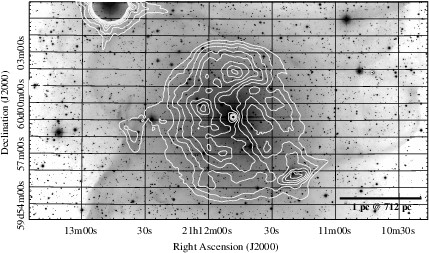}
\caption{\oiii\ emission from the central part of Ou4 vs.\ 22~$\mu$m emission. 
The color scale of the \oiii\ image is logarithmic. 
The contour map is the 22~$\mu$m emission of the $W4$-filter image 
from 89.375 to 92.5~digital numbers (DN) with linear step of 0.446~DN 
(i.e., from 2.471 to 2.557~mJy/arcsec$^2$ with linear step of 0.012~mJy/arcsec$^2$). 
The linear scale is shown in the bottom-right corner.}
\label{figure:oiii_w4}
\medskip
\includegraphics[width=\columnwidth]{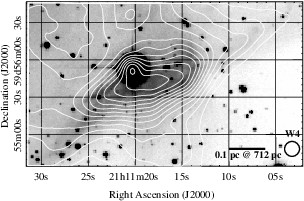}
\caption{Details of the \oiii\ emission of Ou4 near 
the ridge of mid-infrared emission.
The bright star is HD~239597.
The color scale of the \oiii\ image is linear. 
The contour map is the 22~$\mu$m emission from the $W4$-filter image 
from 89.375 to 92.5~digital numbers (DN) with linear step of 0.223~DN 
(i.e., from 2.471 to 2.557~mJy/arcsec$^2$ with linear step of 0.006~mJy/arcsec$^2$). 
The linear scale and the angular resolution of the $W4$-filter image 
($12\arcsec$-FWHM) are shown in the bottom-right corner.}
\label{figure:details_oiii_w4}
\end{figure} 

\subsection{The central part of Ou4: \oiii\ emission vs.\ mid-infrared bubble}
\label{dust}

\subsubsection{Morphology of the mid-infrared bubble}

\begin{figure}[t]
\begin{tabular}{@{}cc@{}}
\includegraphics[width=0.48\columnwidth]{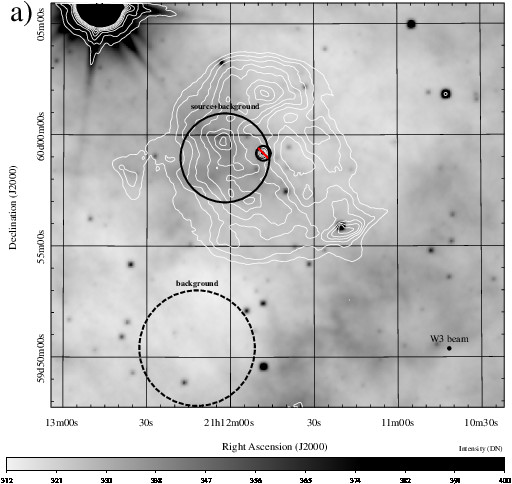}&
\includegraphics[width=0.48\columnwidth]{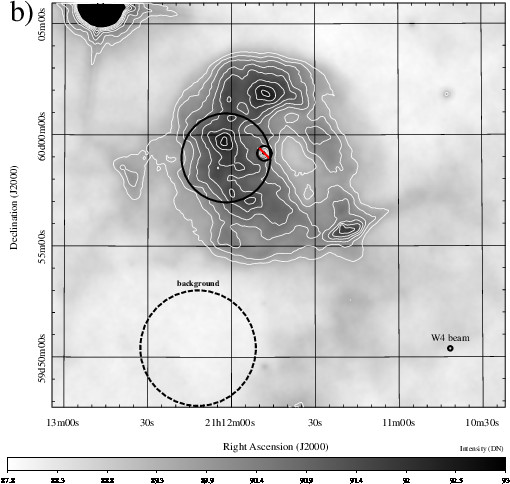}\\
\includegraphics[width=0.48\columnwidth]{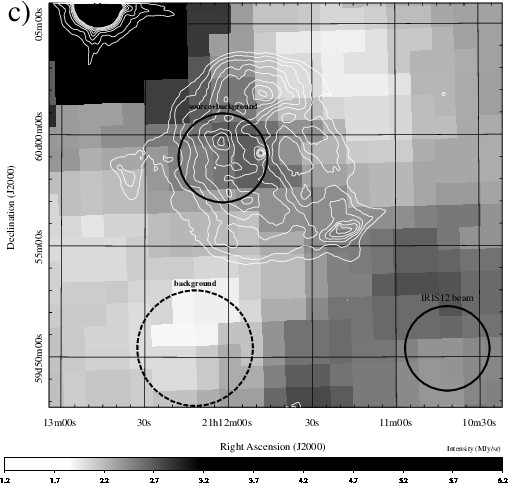}&
\includegraphics[width=0.48\columnwidth]{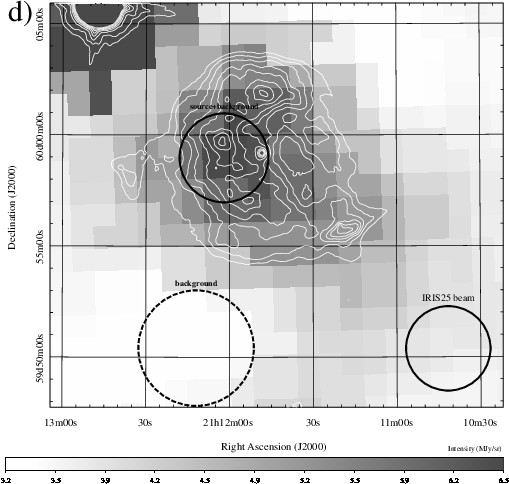}\\
\includegraphics[width=0.48\columnwidth]{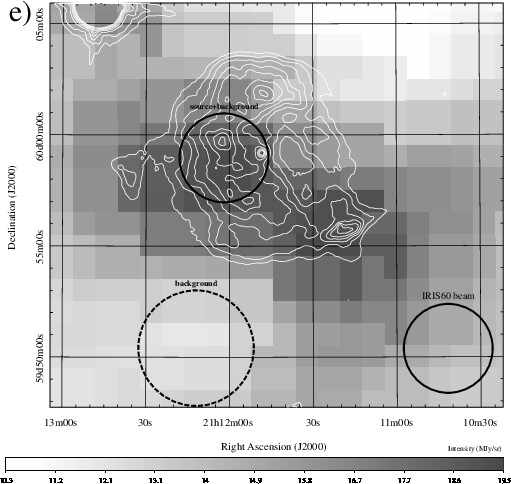} &
\includegraphics[width=0.48\columnwidth]{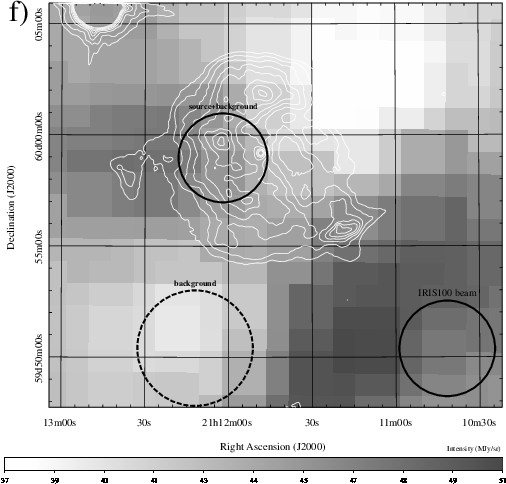}\\
\raisebox{0.22cm}{\includegraphics[width=0.48\columnwidth, trim = 0 1.cm 0 0,clip=true,keepaspectratio]{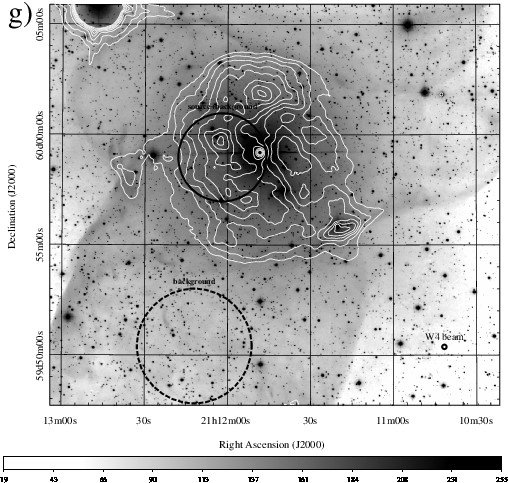}}\\
\end{tabular} 
\caption{Mid- to far-infrared views of the central part of Ou4. 
{\sl a)} WISE 12-$\mu$m ($W3$-filter) image. 
{\sl b)} WISE 22-$\mu$m ($W4$-filter) image.   
{\sl c)} IRIS 12-$\mu$m image. 
{\sl d)} IRIS 25-$\mu$m image.   
{\sl e)} IRIS 60-$\mu$m image. 
{\sl f)} IRIS 100-$\mu$m image. 
{\sl g)} INT \oiii\ image for comparison purpose. 
The contour map is the 22~$\mu$m emission from 89.375 to 92.5~digital numbers 
(DN) with linear step of 0.446~DN (i.e., from 2.471 to 2.557~mJy/arcsec$^2$ with linear step 
of 0.012~mJy/arcsec$^2$). 
The FWHM resolution is plot in the bottom-right corner of each panel.
The two large circles are the regions used for the aperture photometry 
(see text and Fig.~\ref{figure:sed}).
}
\label{figure:view}
\end{figure}

Figure~\ref{figure:oiii_w4} shows a contour map
    of the mid-infrared bubble detected in the WISE $W4$-filter image
    in the central region of Ou4 overlaid on the \oiii\ image.  The
    angular resolution of the $W4$-filter image is $12\arcsec$ (FWHM).
    The central point-like source is the mid-infrared counterpart of
    \hr.  The $5\arcmin$-radius extended emission is asymmetric, with
    a surface brightness which is on average about 3 times higher on
    the eastern-side than on the western-side.  There are two peaks of
    emission, located at $1.8\arcmin$-east and $2.7\arcmin$-north from
    \hr.  The north and south extensions of the mid-infrared bubble
    match the limit of the \oiii\ bubbles (see also the middle-left
    panel of Fig.~\ref{figure:view} for the full range of the $W4$
    intensity).  A weak extension of the infrared bubble towards the
    East of \hr\ corresponds to the base of the \oiii\ streamers.  A
    ridge of mid-infrared emission is located on the south-west border
    of the mid-infrared bubble. Fig.~\ref{figure:details_oiii_w4} is
    an enlargement of this region.  The mid-infrared counterpart of
    the star HD~239597 (K2 spectral type in the Henry Draper catalogue
    and extension) is barely resolved from the ridge emission.  This
    ridge emission corresponds to the \oiii\ filaments that defines
    the southern-limit of the Ou4 bubble.

\begin{figure}[t]
\centering
\includegraphics[width=1.0\columnwidth]{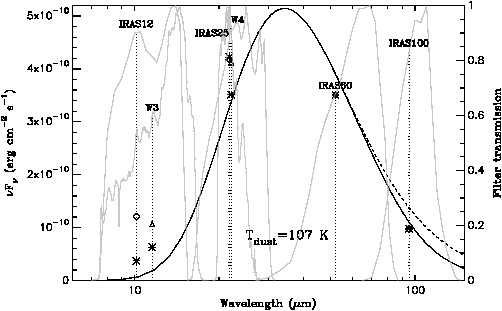}\\
\caption{Spectral energy distribution of the infrared emission in the
  central part of Ou4.  Diamonds and triangles
  are IRIS and WISE data points, respectively, obtained
  from aperture photometry (Fig.~\ref{figure:view}).  
  Grey lines show the transmission profiles of the IRAS and WISE 
  broad-band filters,
  with vertical dotted lines indicating the effective wavelength. 
  The solid line is the dust emission (single-temperature modified black-body model) 
  fitted to the IRIS data points (diamonds) 
  and asterisks are the corresponding modeled fluxes in each filter.
  The dashed line is the corresponding black-body model.
 }
\label{figure:sed}  
\end{figure} 

\subsubsection{Spectral energy distribution of the mid-infrared bubble}

The location of the mid-infrared bubble inside the \oiii\ emission may
suggest line emission from hotter oxygen gas (e.g., \oiv\ at
$\lambda=25.87~\mu$m) rather than continuum emission from dust grains.
Therefore to constrain the nature of the mid-infrared bubble we build
a spectral energy distribution (SED) of the brightest emission peak by
combining WISE ($\lambda_\mathrm{eff}=$11.56 and 22.09~$\mu$m) and
IRAS ($\lambda_\mathrm{eff}=$10.15, 21.73, 51.99,
and 95.30~$\mu$m) photometry.  We use the WISE All-Sky Atlas images
\citep{cutri12} and the {\sl Improved Reprocessing of the IRAS Survey}
\citep[IRIS;][]{miville-deschenes05} which provide us an angular
resolution of $6\farcs5$, $12\arcsec$, $3.8\arcmin$, 
$3.8\arcmin$, $4.0\arcmin$, and $4.3\arcmin$ (FWHM)
at 12, 22, 12, 25, 60, and 100~$\mu$m,
respectively.

Despite the lower angular resolution of IRAS, the east-side of the
mid-infrared bubble is detected in the IRIS~25-$\mu$m image (see the
middle-right panel of Fig.~\ref{figure:view}).  We use a custom {\tt
  IDL} program to perform circular-aperture photometry using a
procedure adapted from {\tt DAOPHOT}.  In all images, the
source$+$background area is centred on ($21^{\mathrm h}12^{\mathrm
  m}02\arcsec$, $+59\degr58\arcmin58\arcsec$; J2000) with a radius of
$2\arcmin$ to match the angular resolution of IRIS images, 
and the background area is centred on ($21^{\mathrm
  h}12^{\mathrm m}12\farcs0$, $+59\degr50\arcmin24\arcsec$ {; J2000}) with a
radius of $2\farcm6$ (Fig.~\ref{figure:view}). We subtract from the
extended-source fluxes in the $W3$, $W4$, IRAS12, and IRASS25 bands the resolved fluxes
of \hr\ in the WISE All-Sky catalog \citep{cutri12}.

The resulting SED (Figure~\ref{figure:sed}) shows significant flux emission 
in all bands, which is consistent with continuum emission from dust grains. 
Assuming that all grains have the same size distribution and composition, 
the dust emission is a single-temperature modified black-body emission, i.e., a black-body 
emission multiplied by the dust emissivity (e.g., \citealt{draine03c}): 
$F_\nu=\Omega \times B_\nu(T_\mathrm{dust})\times [1-\exp\{-\tau_\mathrm{dust}(\nu)\}]$ where 
$\Omega$ is a solid angle, 
$B_\nu(T_\mathrm{dust})$ is the black-body of temperature $T_\mathrm{dust}$, 
and $\tau_\mathrm{dust}(\nu)$ is the optical depth. 
The dust optical depth along the line of sight is defined by: 
$\tau_\mathrm{dust}(\nu)=\kappa_\mathrm{dust}(\nu) \int{\!\rho\,ds}$, 
where $\kappa_\mathrm{dust}(\nu)$ is the dust opacity, and $\rho$ is the dust mass-density. 
We use the dust opacity corresponding to the Milky Way dust with $R_\mathrm{V} = 3.1$, computed 
for the carbonaceous/silicate dust model \citep{weingartner01,draine03}. 
The low foreground-extinction can be neglected.
We convolve $F_\nu$ with the transmission profiles of the broad-band filters\footnote{We use the Filter Profile 
Service of the Virtual Observatory available at 
\href{http://svo2.cab.inta-csic.es/svo/theory/fps3/}{http://svo2.cab.inta-csic.es/svo/theory/fps3/}\,.}.
We fit {\sl only} the IRIS data points to mitigate the variation of angular resolution with wavelength. 
Our best fit\footnote{The two other physical parameters 
are: the diameter of the black-body region ($\equiv2d\sqrt{\Omega/\pi}$) of 89~AU for a distance, 
$d$, of 712~pc; 
and the dust surfacic mass ($\equiv\int{\!\rho\,ds}$) of 0.054~g\,cm$^{-2}$.} is obtained 
for a dust temperature of 107~K.
The decrease of dust emissivity with wavelength affects only the fluxes 
in the $IRAS60$ and $IRAS100$ bands (see dashed line in Figure~\ref{figure:sed}).
The excess of emission in the $IRAS12$ and $W3$ bands may be due to additional 
emission from the background cloud.
The excess of emission observed in the $W4$ band compared to the model prediction 
is likely due to the better angular 
resolution of WISE compared to IRIS. 
Our model predicts a maximum of dust emission around 34~$\mu$m.

In conclusion, the overall SED is typical of continuum emission from
hot dust grains.

\section{Discussion and conclusions}
\label{discussion}

Summarising, one of the main results of this study is that the tip of
the south lobe of Ou4 is shock ionised. This is clearly indicated by
the spatial distribution of the atomic emission, with the higher
ionisation species outwards and the lower excitation, cooling
post-shock gas inward, as well as by the observed line ratios and gas
physical conditions.

The fact that this region is ionised by collisions rather than by
photons, does not require that the central source at the origin of the
outflow has the high temperature which would be needed to excite,
e.g., O$^{2+}$. Removing this constraint on the temperature of the
central source may weaken the hypothesis that Ou4 is a PN.  However,
we do not have information about the ionisation mechanism of the inner
regions of Ou4. It may well be that, in addition to shocks in the
outermost regions outflow, a small
photoioinised core is present, such as in the case of the PN KjPn~8
\citep{lopez95}.

We estimate the probability of an apparent association between
  \hii\ regions and PNe by cross-correlating the WISE catalogue of
  \hii\ region \citep{anderson14}, which provides \hii\ angular radii
  from $6\arcsec$ to $1\fdg6$, with the catalogue of Galactic PNe of
  \cite{kohoutek01}.  We identify 29 on 1510 PNe (i.e., 1.6\%)
  that are located inside \hii\ regions in projection. The mean and
  standard deviation of their Galactic latitudes is $-0\fdg2$ and
  $1\fdg5$, respectively, with a maximum angular distance from the
  Galactic plane of $3\fdg9$.  Therefore, lucky association of
  \hii\ regions and PNe only occurs close to the Galactic plane
  due to the high spatial-density of both kind of objects in this
  region of the sky. Since the Galactic latitude of \sh\ is high
  ($b\approx8\degr$), a fortuitous alignment with a PN appears as
  unlikely.
 
Given the apparent location of Ou4 in the sky, aligned with the young
stellar cluster at the centre of the \hii\ region \sh\ and the
  striking correspondence between the 22~$\mu$m and \oiii\ emission,
  it is reasonable to suppose that Ou4 is an outflow launched some
  90,000~yr ago from the massive triple system \hr. The location of
the outflow, its radial velocities and extinction values are all
consistent with such an hypothesis.  Further support might come from
the fact that the south bow-shock has an enhanced brightness in
correspondence with a bright portion of \sh, as expected if it were
impinging on a denser zone at the border of the \hii\ region.


For comparison purpose, we estimate the kinetic energy of the Ou4
outflow. With lobe size of 2.5~pc $\times$ 14.4~pc in extent, for a
mean \hii\ density of 1.7~cm$^{-3}$ and $\mu=0.7$, the displaced mass
is $M_\mathrm{disp}\approx 2.1~M_\odot$.  With a shock propagating at
$\approx$100~\kms\ the kinetic energy of the outflow is
$K(\mathrm{outflow})\approx4\times10^{47}$~ergs, i.e., much lower than
SNe.  If the bipolar-cavity was created by an episodic, collimated,
bipolar jet/wind propagating at 2,500
\kms\ \citep[e.g.,][]{steffen98}, the conservation of the
kinetic-energy would imply a mass ejected by the driving source of
about $M_\mathrm{ejec}\approx 0.003~M_\odot$.  With a dynamical-time
of 88,000~yr, the required mass-loss rate would be therefore
$3.8\times10^{-8}~M_\odot$~yr$^{-1}$.  This value can be compared with
the mass-loss rate from the radiatively-driven winds of the central
massive stars.  From the physical parameters of the stellar components
Aa, Ab, and B of \hr\ in Sect.~\ref{distance} we estimate using the
theoretical recipe of \cite{vink00} the mass-loss rates of
$\approx2.5\times10^{-8}$, $\approx0.6\times10^{-8}$, and
$\approx1.6\times10^{-8}$~$M_\odot$~yr$^{-1}$ with terminal velocities
of 2515, 2508, and 2517~\kms, respectively.  Therefore, the two-most
massive stars of \hr\ with combined mass-loss rate of
$4.1\times10^{-8}$~$M_\odot$~yr$^{-1}$ and high terminal velocity can
easily provide the required kinetic energy to drive Ou4.  The larger
separation of component Aa and B may explain the episodic phenomenon,
whereas the close Ab companion could play a role in the outflow
collimation.

In the alternative scenario that Ou4 is a PN, its overall morphology
would likely be the result of the expansion of a fast collimated wind
from a yet unidentified central source through a relatively dense
circumstellar medium. If the cooling time of the shocked material
(fast wind and/or ambient medium) is long enough, the fast collimated
wind can inflate the observed thin lobes. The morphology of these
lobes depends on several factors, mainly the opening angle of the fast
outflow \citep{soker04}. In addition, as discussed by \cite{sahai99}
for the pre-PN Hen~401, the ambient density should decrease
significantly with distance from the source in order to produce the
cylindrical nebular morphology of the southern lobe.
The presence of bow-shocks and the ``bowl'' shape of the base of the
southern lobe star are also consistent with models considering the
inflation of bipolar lobes by fast collimated outflows seen nearly
edge-on \citep{soker02,balick13}.  

The study of this kind of collimated outflows in PNe and related
objects often suggests an {\sl eruptive} nature of the phenomenon, and
that the central source is a binary system.  The possibility that the
outflow of Ou4 is produced in an outburst powered by mass accretion in
a binary system, leading to a phenomenon such as an
intermediate-luminosity optical transient (ILOT) as proposed for
e.g. KjPn~8 \citep{boumis13}, other bipolar PNe \citep{soker12}, or
for massive ($M_{ZAMS}$$\sim$6--10~$M_\odot$) carbon-rich AGB,
super-AGB, or post-AGB stars \citep{prieto09}, is an appealing one.

Concluding, it is clear that the stellar source at the origin of the
Ou4 outflow should be better constrained.  A crucial information in
this respect is its distance. In the future, we aim at a precise
measurement of the proper motions of the outflow which, combined with
the line-of-sight velocities and kinematical modelling, would provide
a sound distance determination.  A more sensitive imaging of the
vicinity of the bright star \hr\ with narrower filters centred on
emission lines would be also valuable to detect the origin of the
outflow.  In addition, extension of this study to other wavelength
domains is planned to better constraint the nature of this unique
giant outflow.

\begin{acknowledgements}
We thank the referee John Bally and the editor Malcolm Walmsley 
for many useful suggestions that helped to improve our manuscript.  We
are grateful to the time allocation committee (CAT) for awarding us
IAC Director Discretionary Time at the WHT and INT. The higher
resolution spectra was secured during ING service time. We thank the
ING staff, and in particular Javier Mendez, Raine Karjalainen, and the
ING students for imaging attempted on an additional night, which could
not be used due to the strong moonlight which prevented detection of
the faint nebula. RLMC acknowledges funding from the Spanish
AYA2007-66804 and AYA2012-35330 grants.  We thank St{\'e}phane Zoll
for his \oiii\ image of Ou4.  Finally, we are very grateful to Gabriel
Perez at the IAC for the careful edition of the colour image in
Fig.~\ref{F-wholeneb}.  This research has made use of Aladin, and of
the SIMBAD database, operated at CDS, Strasbourg, France. This
publication makes use of data products from the Wide-field Infrared
Survey Explorer, which is a joint project of the University of
California, Los Angeles, and the Jet Propulsion Laboratory/California
Institute of Technology, funded by the National Aeronautics and Space
Administration.
\end{acknowledgements}


\bibliographystyle{aa}
\bibliography{ou4_low_res.bbl}

\end{document}